\documentclass[preprint,showpacs]{revtex4}
\usepackage{epsfig}

\newcommand{\ba}{\begin{eqnarray}}
\newcommand{\ea}{\end{eqnarray}}

\begin{document}
\title{Unexpected residual regularities in liquid drop mass calculations}

\author{Jorge G. Hirsch,  Alejandro Frank, and V\'\i ctor Vel\'azquez}
\affiliation{Instituto de Ciencias Nucleares, 
Universidad Nacional Aut\'onoma de M\'exico, \\
Apartado Postal 70-543, 04510 M\'exico, D.F., M\'exico\\
E-mail: hirsch@nuclecu.unam.mx, frank@nuclecu.unam.mx, vic@nuclecu.unam.mx}

\begin{abstract}
A systematic study of correlations in the chart of nuclear masses 
calculated using the Finite Range Droplet Model of M\"oller et al. is presented.
It is shown that the differences between the calculated and measured masses 
have a well defined oscillatory component as function of the proton and neutron numbers, 
which can be removed with an appropriate fit, reducing significantly the error width, and 
concentrating the error distribution on a single peak around zero. 
The presence of this regular residual correlations suggests that the Strutinsky
method of including microscopic fluctuations in nuclear masses could be improved.
\end{abstract}

\maketitle

\section{Introduction}

The prediction of nuclear masses is of fundamental importance for
a complete understanding of the nuclear processes that power the Sun,
and the synthesis and relative abundances of the elements \cite{Rol88}.
M\"oller at al. \cite{Moll95}, Duflo and
Zuker \cite{Duf94}, Goriely et al. \cite{Gor01}, among many others, have developed mass formulae 
that calculate and predict the masses (and often other properties)
of as many as 8979 nucleotides.
There is a permanent search for better parameterizations that decrease the
difference with the experimental masses and produce reliable predictions for unstable 
nuclei.

The atomic mass excesses have been tabulated for 8979 nuclei ranging from
$^{16}$O to A=339 in \cite{Moll95}, calculated with a Finite Range Droplet 
macroscopic Model and a folded-Yukawa single-particle microscopic model,
FRDM for short.
The microscopic sector includes a Lipkin-Nogami calculation of pairing gaps besides
Strutinsky shell-corrections. Ground-state energies are minimized with respect
to shape degrees of freedom. 
Only 9 constants are adjusted to the ground-state
masses of 1654 nuclei, with a mass model error of 0.669 MeV in the entire
region of nuclei ranging from $^{16}$O to $^{263}$106. 

The average mass error is found to
gradually increase in a systematic way as the lighter region is approached. 
This behavior was related in \cite{Moll95} with the reliability of
the Strutinsky method for the folded-Yukawa single-particle method, which is
expected to be less accurate for light nuclei because the smooth, average quantities
are less accurately determined from the relatively few levels occurring in light nuclei.

\begin{figure}[h]  
  \begin{center}
    \leavevmode 
\vskip -.4cm
     \psfig{file=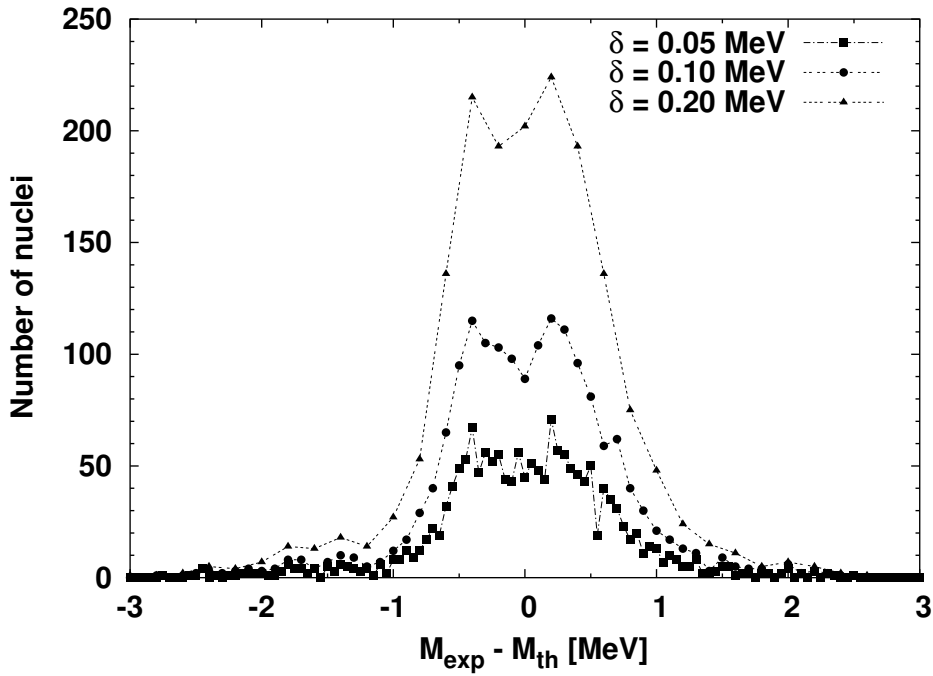,angle=270 , width=13.0cm}
\vskip .2cm
  \caption{Distribution of nuclei with a given mass error $M_{exp}-M_{th}$ for three
histogram intervals.}
    \label{doublepeak}
  \end{center}
\end{figure}

The main motivation for the present work arose from the detailed study of the nuclear mass error distribution shown in Fig. \ref{doublepeak}, 
which displays the number of nuclei within a given interval of FRDM mass errors  \cite{Moll95},
for the 1654 isotopes with measured masses in 1995. The symmetrical
distribution with two peaks around the origin was completely  unexpected. 
In this article we study the regularities in the differences between measured and
calculated masses, showing that they are closely related to the presence of this
double peak. 

Fig. \ref{doublepeak} shows the FRDM distribution of nuclei \cite{Moll95}
for different intervals of mass errors. The intermediate curve (diamonds) 
corresponds to a mass error interval of $0.1$ MeV where the double peak
 is quite evident.
Enlarging the mass error interval to $0.2$ MeV (triangles) softens the curvature 
but the presence of the double peak is still very clear. In
the opposite direction, an interval of $0.05$ MeV (squares) produces larger
 fluctuations and some apparent interference, but it is evident that the
distribution still displays the peaks. We can thus conclude that 
the effect is real and not an artifact of a particular distribution interval. 

We carried out additional tests to verify the persistence of the double peak,
including only nuclei with A $>$ 40 and A $>$ 60, normalizing the distribution
according to the rule $A^{-1/3}$, and including the experimental errors. 
In all circumstances the double peak remained \cite{Vel03}.

\begin{figure}[h]  
  \begin{center}
    \leavevmode 
    \psfig{file=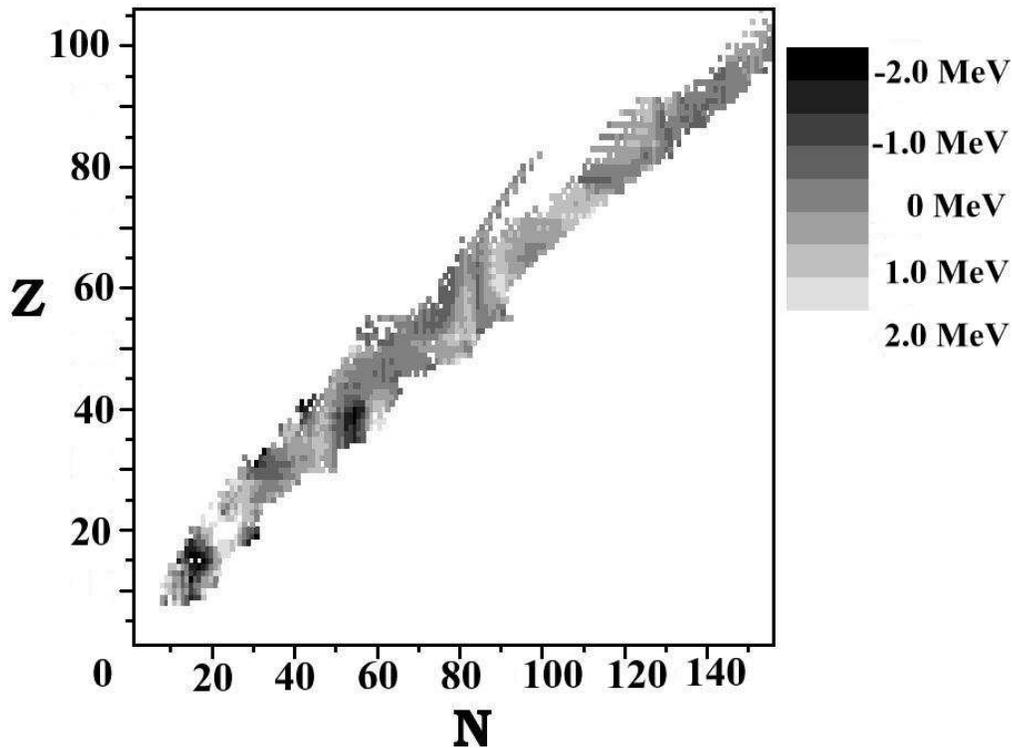, width=14.0cm}
\vskip -.5cm
    \caption{Distribution of $\Delta M(N,Z)$ in the $N-Z$ plane.
The gray mass scale is given on the right.}
    \label{dif-mn}
\vskip -.5cm
  \end{center}
\end{figure}

Figure \ref{dif-mn} contains the distribution of FRDM mass differences \cite{Moll95}
\begin{equation}
\Delta M(N,Z) \equiv M_{exp}(N,Z) - M_{th}(N,Z),
\end{equation}  
in the proton number (N) - neutron number (Z) space. We 
can see large domains with a similar error (each  tone is
associated to the magnitude of the error). It is remarkable that  
very well defined correlated areas of the same gray tone exist, which
is a clear indication of remaining systematics and correlation.

Both the distribution of mass errors in the N-Z plane and the presence of the double peak
in the error distribution suggest the presence of important correlations which
are not taken into account in the FRDM mass formula. In the next sections
these correlations are quantified and parametrized.

\section{Regularities in the mass errors}

In order to visualize the oscillatory patterns suggested in Fig. \ref{dif-mn}, 
different cuts were performed along selected directions on the N-Z plane.  
Given the large number of chains which can be studied, we have selected those
with the largest number of nuclei with measured mass.
For each cut a Fourier analysis was performed, and the squared amplitudes are plotted as a
function of the frequencies on the right hand side of each figure.
The discrete Fourier transforms $F_k$ are calculated as
\begin{equation}
F_k = {\frac {1} {\sqrt{N}}} \sum_j {\frac {\Delta M(j)}{\gamma}} \, \hbox{exp}
\left({\frac {-2 \pi i j k} {N}}\right) ,
\end{equation}
where N is the number of mass differences $\Delta M$ in a given series.
The parameter $\gamma$ makes $F_k$ dimensionless. Given that it only affects 
the global scale of the Fourier amplitudes, we made the simple selection
$\gamma = 1$ MeV.
The Fourier amplitudes are plotted as functions of the frequency $f = k/N$.
 The presence of a few frequencies with notorious large
components underlines the existence of an oscillatory behavior of the mass errors.

\subsection{Fixed N or Z}

\begin{figure}[h]  
  \begin{center}
    \leavevmode 
    \psfig{file=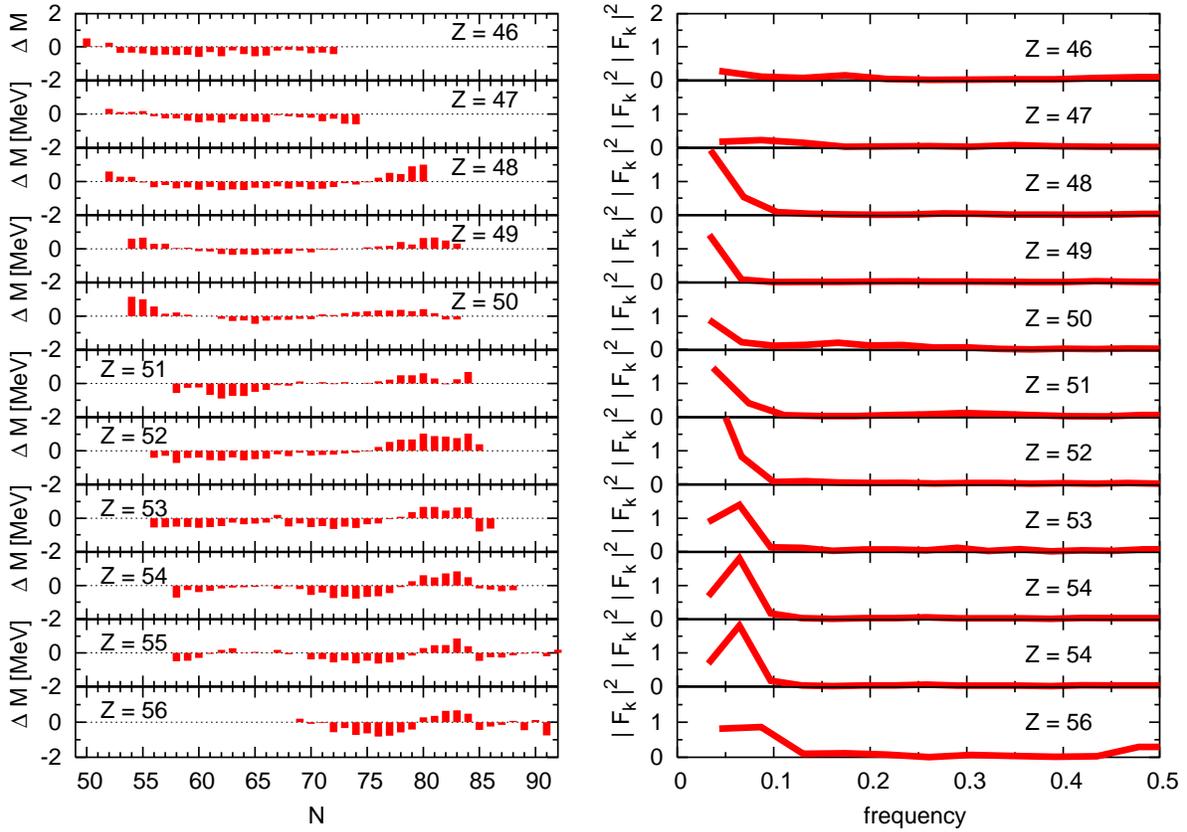,angle=270,width= 16.cm}
 \end{center}
\vskip -.9cm
\caption{Mass errors in the isotope chains Z=46 to 56, and their Fourier analysis.}
\label{z46}
\end{figure}

\begin{figure}[h]  
  \begin{center}
    \leavevmode 
    \psfig{file=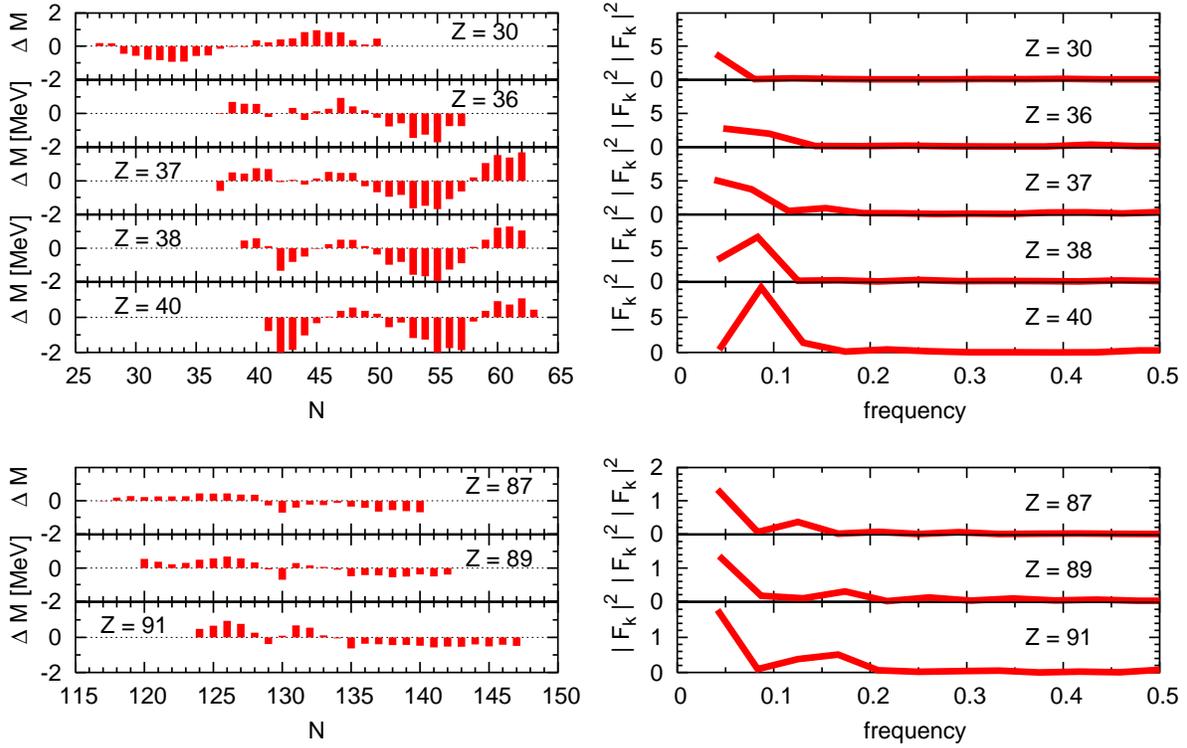,angle=270,width= 16.cm}
 \end{center}
\vskip -.9cm
\caption{Mass errors in the isotope chains with Z = 30,36, 37, 38, 40, 87, 89, 91, 
and their Fourier analysis.}
\label{z30}
\end{figure}

We start our analysis for fixed N or Z, i.e. we selected different chains of isotopes and 
isotones. Those isotopic chains with 20 or more nuclei with measured masses are
presented in Fig. \ref{z46} and \ref{z30}.
Until 1995, the element which had most 
isotopes with measured masses was Cs (Z=55) with 34. 

Fig. \ref{z46} displays the mass errors for the isotope 
chains Z = 46 to 56, and 
their Fourier analysis, nearly all exhibiting a clear peak for the low frequency 
$f \approx 1/20 = 0.05$.

Fig. \ref{z30} displays the mass errors for the isotope chains 
Z = 30,36, 37, 38, 40, 87, 89, 91,  and 
their Fourier analysis.

\begin{figure}[h]  
  \begin{center}
    \leavevmode 
    \psfig{file=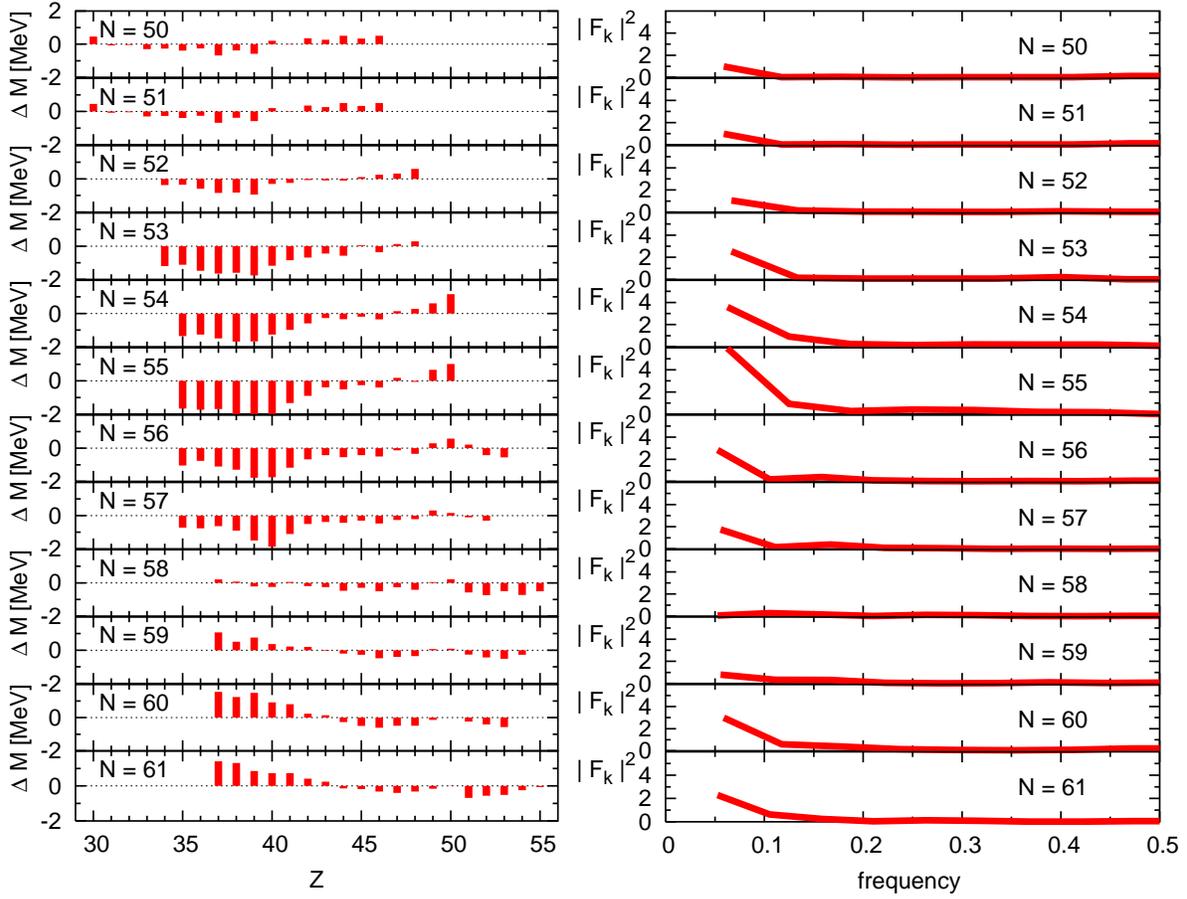,angle=270,width= 16.cm}
 \end{center}
\vskip -.9cm
\caption{Mass errors in the isotone chains N=50 to 61, and their Fourier analysis.}
\label{n50}
\end{figure}

A similar analysis was performed for fixed N, i.e. isotonic chains, which have
up to 30 nuclei with measured masses.
Fig. \ref{n50} displays the mass errors for the isotone chains N=50 to 61, and 
their Fourier analysis.

\begin{figure}[h]  
  \begin{center}
    \leavevmode 
    \psfig{file=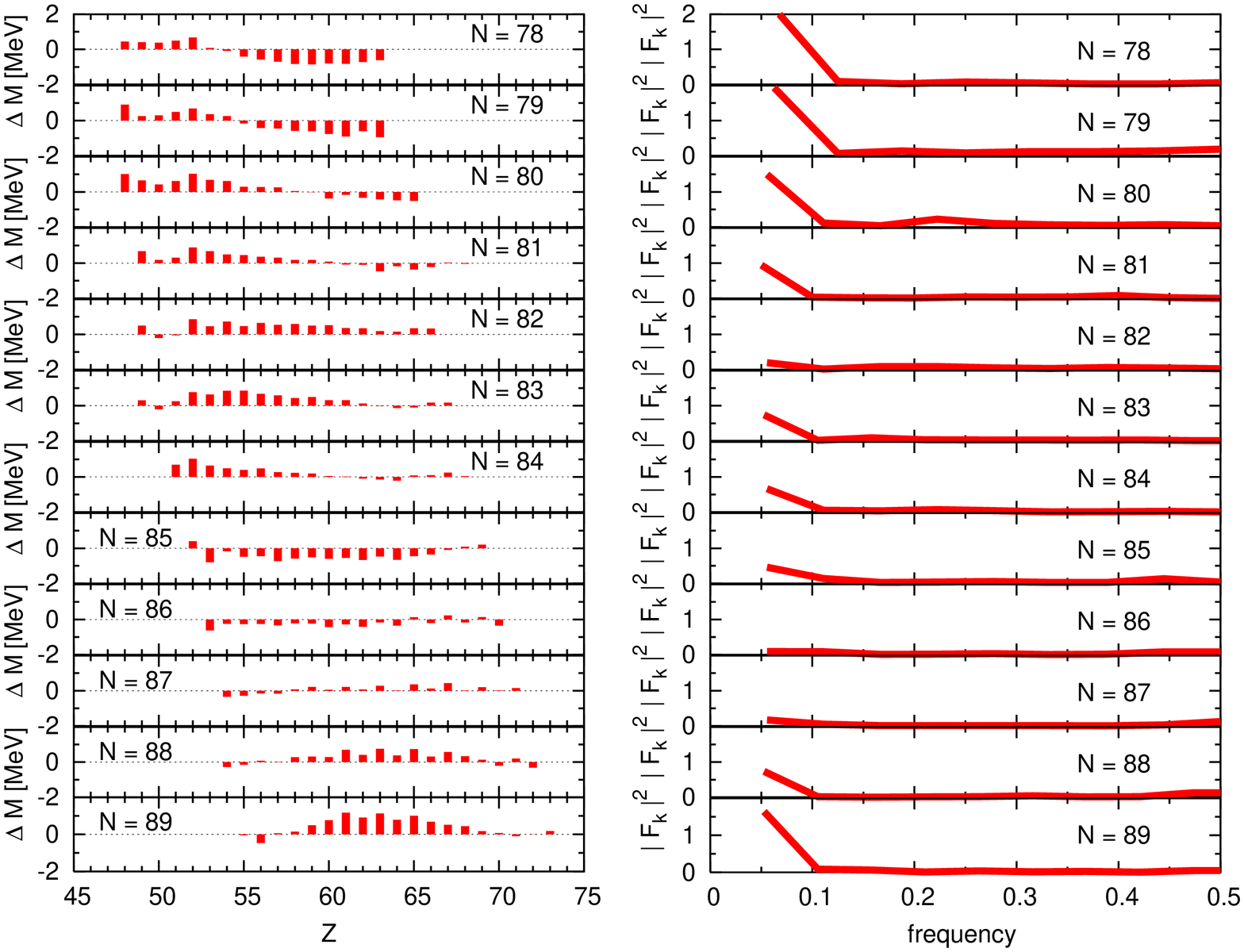,angle=270,width= 16.cm}
 \end{center}
\vskip -.9cm
\caption{Mass errors in the isotone chains N=78 to 89, and their Fourier analysis.}
\label{n78}
\end{figure}

Fig. \ref{n78} displays the mass errors for the isotone chains N=78 to 89, and 
their Fourier analysis.

Many isotopic chains exhibit similar oscillatory patterns in the mass errors. 
However, a Fourier decomposition for these cases is crude because there are only a few points, 
typically between 15 and 30, in each chain. 

\subsection{Fixed A or T$_z$}

To map the mass error data in term of variables with the maximum possible number
of nuclei along each chain, the following transformation is employed
\begin{equation}
\tilde A = Int[\sqrt{2} \,( N \,sin \theta \,+\, Z\, cos \theta )], ~~~~
\tilde T_z = Int[\sqrt{2}\, ( N \, cos \theta \,-\, Z \,sin\theta )].
\label{a-tz}
\end{equation}
Both $\tilde A$ and $\tilde T_z$ are, by construction, integer numbers.
To avoid introducing artificial noise, the data are {\em softened} by the 
interpolation of mass errors for {\em unphysical} values  of $\tilde T_z, \tilde A$,
i.e. those with $\tilde T_z$ even and $\tilde A$ odd, or viceversa. This process is necessary
to eliminate the large number of zeros which are induced by the transformation,
which create artificial high frequency noise in the data.

\begin{figure}[h]  
  \begin{center}
    \psfig{file=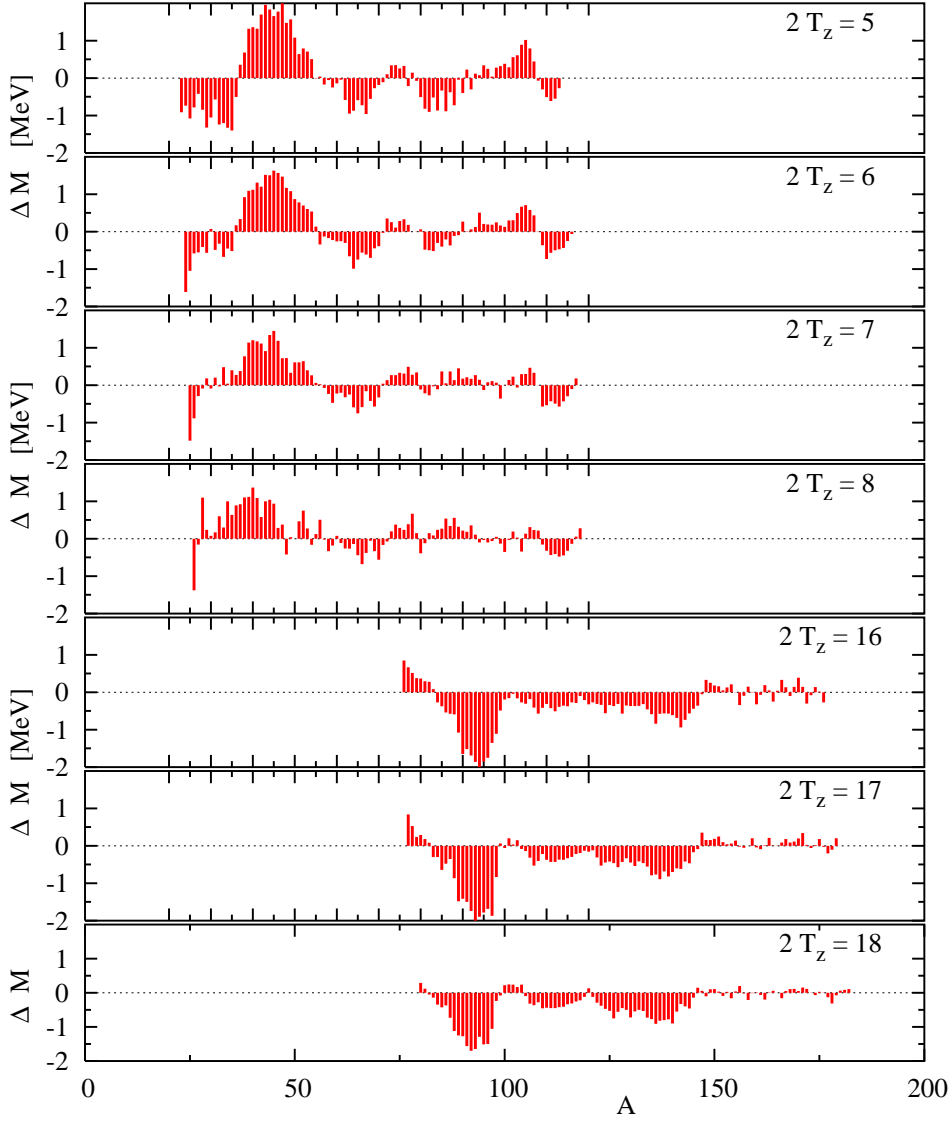, width=13.0cm}
  \end{center}
\vskip -3.cm
\caption{Mass differences as functions of $A$, for seven $T_z$ values.} 
\label{difmass-45}
\end{figure}

Definition (\ref{a-tz}) has the advantage that, for $\theta = 45 ^\circ$, 
$\tilde A = A, \tilde T_z = 2 T_z$. There are seven values of $T_z$ which
have more than 40 nuclei with measured masses. They are $2 T_z = 5, 6, 7, 8, 16, 17 ,18$.
Their mass differences $\Delta M$ are plotted as a function of the mass number $A$ in
the seven inserts shown in Fig. \ref{difmass-45}. The clustering of negative and
positive errors is evident, again exhibiting the presence of
residual correlations between the mass errors and the atomic and neutron numbers
(Z, N), or, equivalently, with the isospin projection and the mass numbers.

\begin{figure}[h]  
  \begin{center}
    \psfig{file=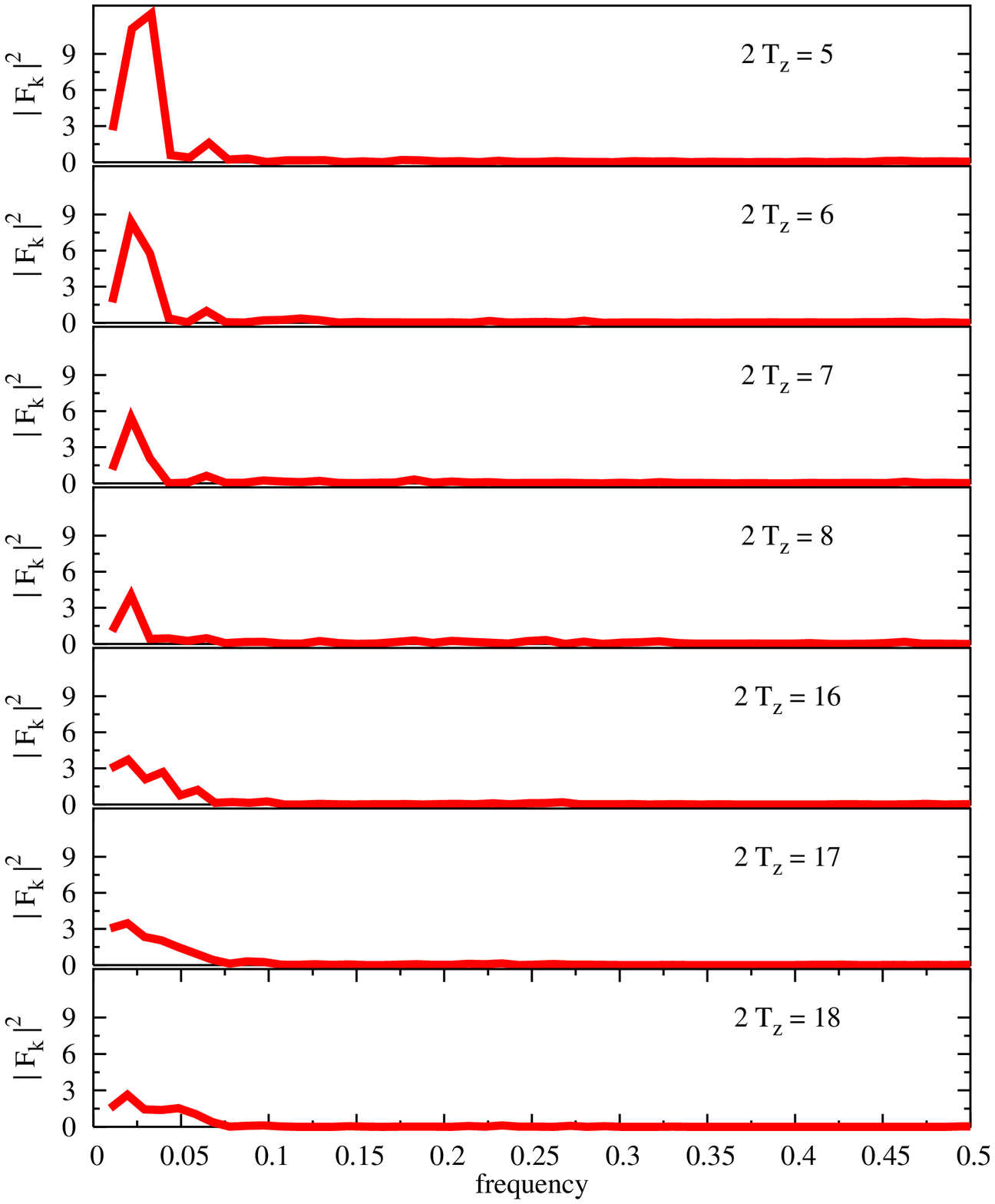, width=13.0cm}
  \end{center}
\vskip -3.cm
\caption{Squared amplitudes of the Fourier transforms of the mass differences, 
plotted as functions of $A$, for seven $T_z$ values.} 
\label{fourier-45}
\end{figure}

The squared amplitudes of the Fourier transforms of the mass differences are shown
in Fig. \ref{fourier-45}. For the first four isotopic chains ($2 T_z= 5, 6, 7, 8$)
there is a prominent peak at $f \approx 0.033$, i.e. a period $\Delta A \approx 30$, which
is about one third the size of each set of nuclei. The remaining three chains exhibit also a
peak at low frequencies, but lower and wider.

\begin{figure}[h]  
  \begin{center}
    \psfig{file=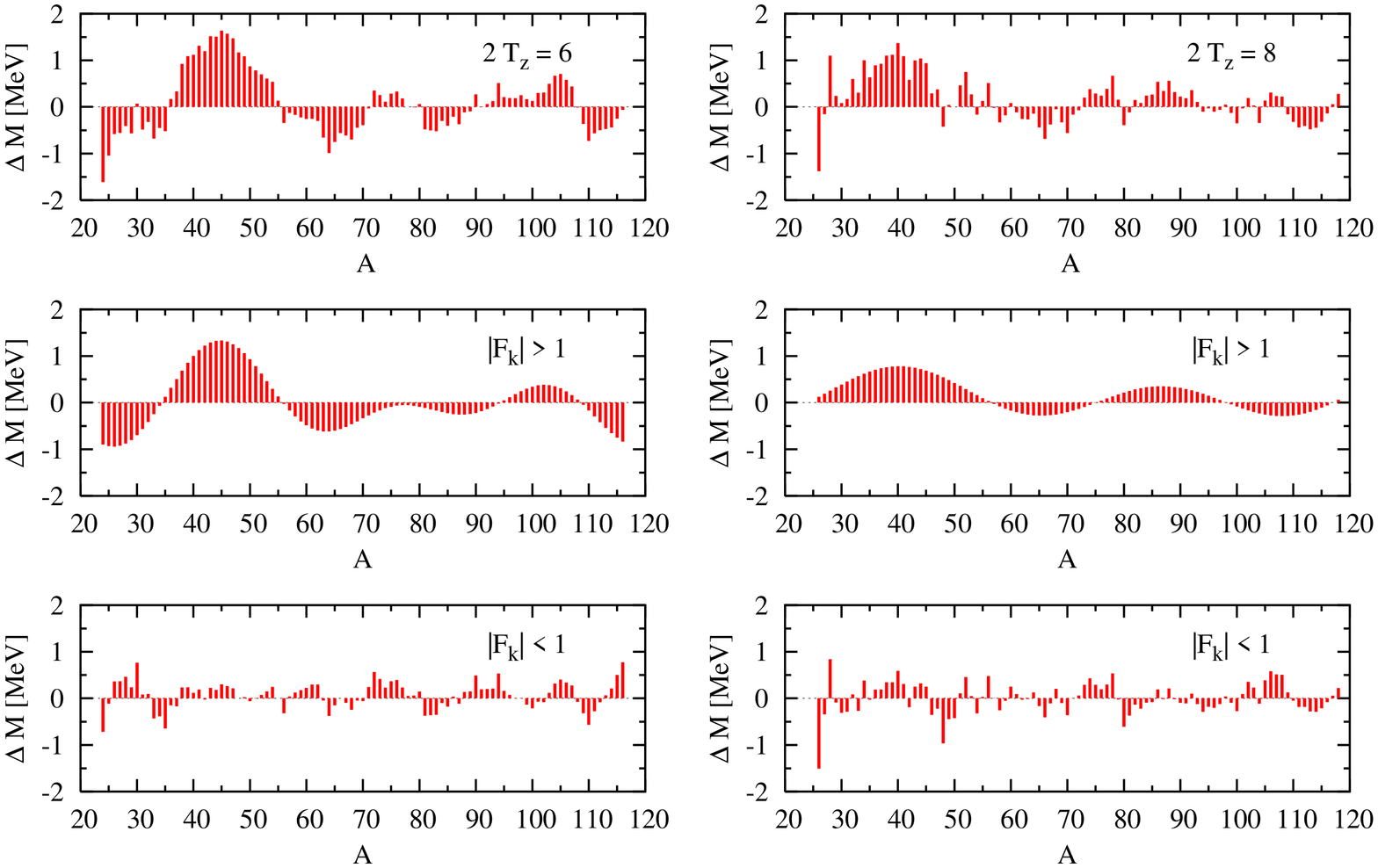, angle=270, width=16.0cm}
  \end{center}
\vskip -.9cm
\caption{Mass differences as functions of $A$, for $2 T_z = 6$ (left), and 8 (right).
The middle panels show the distribution for only the few frequencies with Fourier amplitudes
larger than 1, the lower panels the distribution for all the remaining frequencies.} 
\label{difmass45-tz6}
\end{figure}

In order to underline the relevance of the lower frequencies in the mass error distribution,
in Fig. \ref{difmass45-tz6} the mass errors are plotted as a function of A, for $2 T_z = 6$ 
(left), and 8 (right), 
in the top panel. The middle panel exhibits the distribution obtained including only the three
low frequencies: $f = 0.011, 0.022, 0.033$. The second is the most important, with a period
$\Delta A \approx 46$.

\subsection{Rotated $\tilde A$ and $\tilde T_z$}

 We found that the best orientation, in order to have
as many isotopes as possible with the same $\tilde T_z $, is $\theta = 56 ^\circ$.
With this transformation, there are 174 isotopes with $\tilde T_z = 0$.

\begin{figure}[h]  
  \begin{center}
    \psfig{file=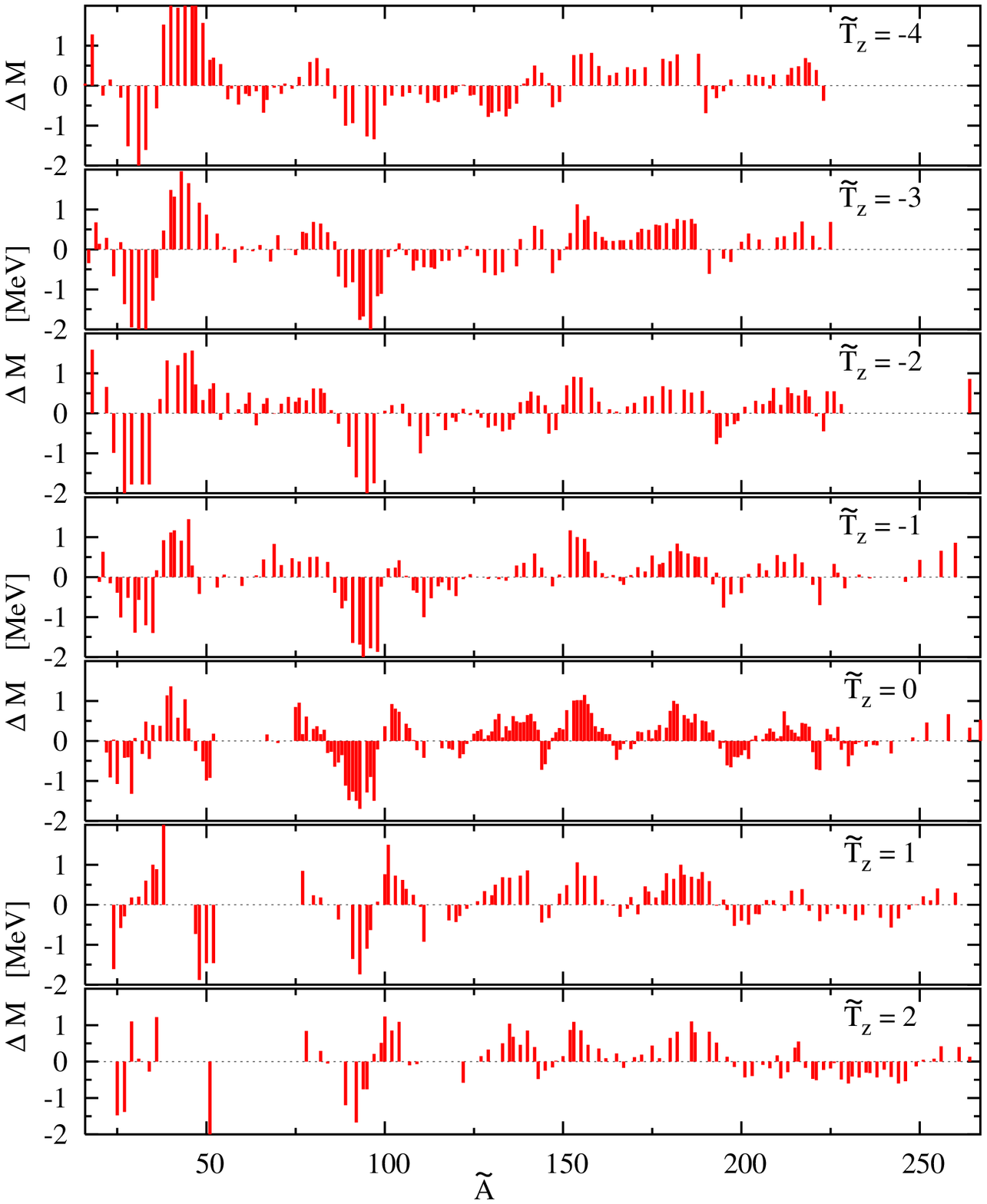, width=13.0cm}
  \end{center}
\vskip -3.cm
\caption{Mass differences as functions of $\tilde A$, for seven $\tilde T_z$ values.} 
\label{difmas-tz}
\end{figure}
Figure \ref{difmas-tz} displays the mass errors for 7 values of $\tilde T_z$, from
 $\tilde T_z = -4$ to 2. The regularities seen in Fig. \ref{dif-mn} as regions
with the same gray tone are seen here in the different plots, as groupings of nuclei
with similar positive or negative mass differences, for the same $\tilde A$ region.
Besides the two large groups with positive and negative mass errors below 
$\tilde A=50$, there are evident regions with negative errors close to
$\tilde A=100$, and with positive mass differences for $150< \tilde A <200$.

\begin{figure}[h]  
  \begin{center}
    \psfig{file=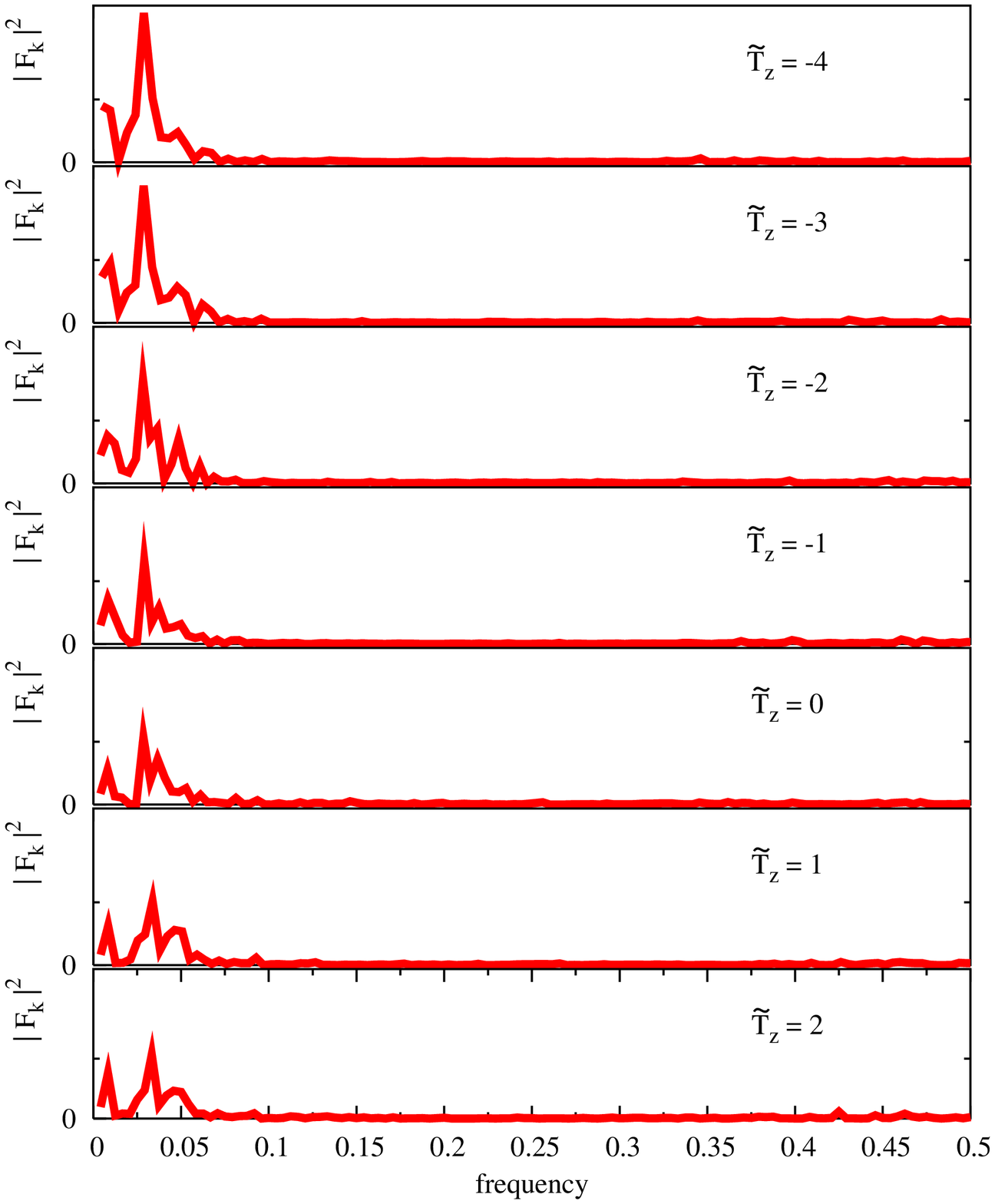, width=13.0cm}
  \end{center}
\vskip -2.cm
\caption{Squared amplitudes of the Fourier transforms of the mass differences, 
plotted as functions of the frequency, for seven $\tilde T_z$ values.} 
\label{four-difmas56-tz}
\end{figure}
A Fourier analysis of the previous results is presented in Fig. \ref{four-difmas56-tz}.
It is clear that for all the chains a few low frequencies again
dominate the spectrum. In some chains
there are also some higher frequencies which seem to be relevant, while for $\tilde T_z = 0$
their contribution is small. These frequencies, close to $f = 0.5$, are associated to oscillations
with period 2, i.e. strong fluctuations between one nucleus and its closest
neighbors.

\begin{figure}[h]  
  \begin{center}
    \psfig{file=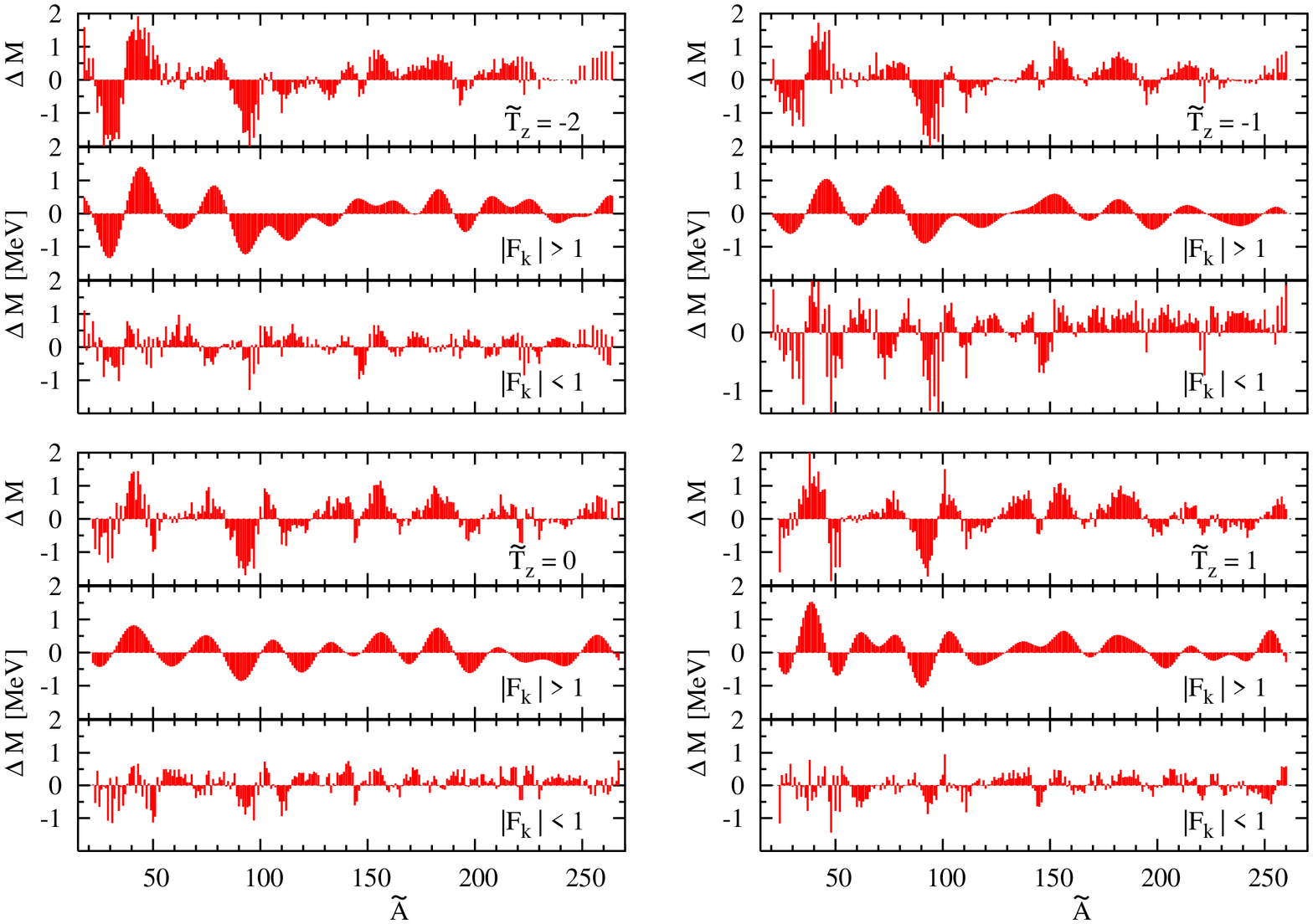, angle=270, width=16.0cm}
  \end{center}
%\vskip -.9cm
\caption{Mass differences as functions of $\tilde A$, for $\tilde T_z = -2$ (upper left), 
-1 (upper right), 0 (lower left) and 1 (lower right).
The middle panels show the distribution for only the few frequencies with Fourier amplitudes
larger than 1, the lower panels the distribution for all the remaining frequencies.} 
\label{difmass56-tz-2}
\end{figure}
In Fig. \ref{difmass56-tz-2} the contribution of very few frequencies to the mass differences,
for the chains with $\tilde T_z = -2, -1, 0$ and $1$ are presented. The upper panels show
the full sequences of mass differences, the medium panels the contribution of the four 
frequencies
with the largest amplitudes, and the lower panels the remaining mass errors. In all cases
it is apparent that the regularities are quite significant, and that the substraction of these systematic errors would lead to important improvements in mass calculations.

\subsection{Bustrofedon: all data aligned}

\begin{figure}[h]  
  \begin{center}
    \psfig{file=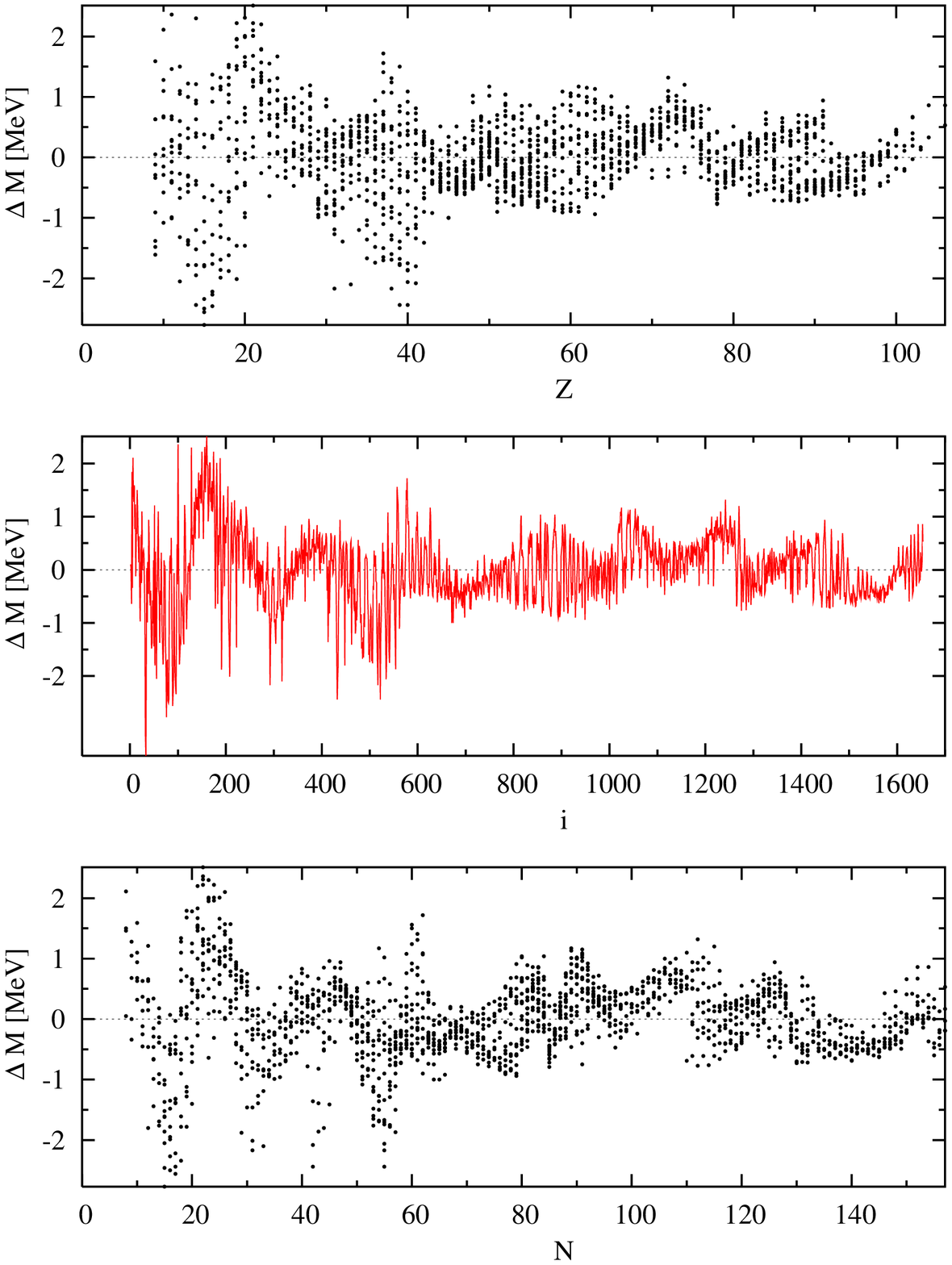, width=13.0cm}%,angle=270}
  \end{center}
\vskip -2.cm
\caption{Mass differences plotted as function of Z (top), N (bottom), and of an ordered list
(middle)} 
\label{difmas-l}
\end{figure}

\begin{figure}[h]  
  \begin{center}
    \psfig{file=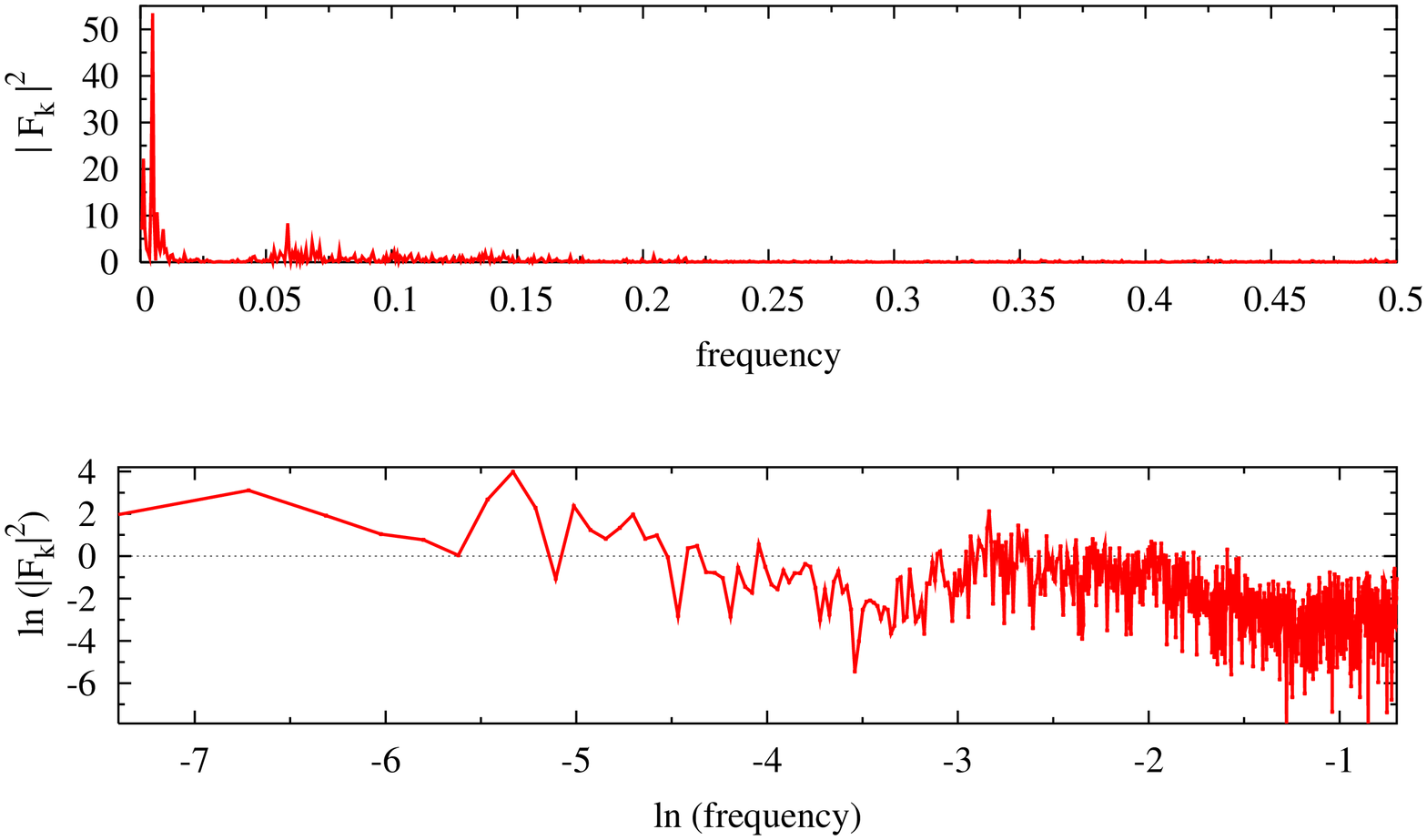, width=11.0cm,angle=270}
  \end{center}
%\vskip -3.cm
\caption{Squared amplitudes of the Fourier transforms of the mass differences, 
plotted as functions of the frequency (top). Power spectrum: natural logarithm of squared amplitudes of 
the Fourier transforms of the mass differences (bottom).} 
\label{fourier-l}
\end{figure}

Plotting the mass differences for different Z, Fig. \ref{difmas-l} top, and
for different N,  Fig. \ref{difmas-l} bottom, is very common in mass calculations.
Both plots exhibit some degree of structure. In this way we obtain a plot of mass 
differences as a function of Z, with all the isotopes
plotted along the same vertical line, see Fig. \ref{difmas-l}.
The difficulty in quantifying these
regularities lies in the simple fact that the are many nuclei with a given N or Z.
For this reason in the previous subsections we have analyzed the data using different cuts.
Another way to organize the FRDM mass errors for the 1654 nuclei with measured masses
is to order them in a single list, numbered in increasing order. To avoid jumps,
we have ordered the isotopes along a $\beta{}o\upsilon\tau\rho{}o\phi\eta\delta\acute{o}\nu$ 
(bustrofedon) line, which literally means "in the way the ox ploughs".
Nuclei were ordered in increasing mass order. For a given even A, they were
accommodated following the increase in N-Z, and those
nuclei with odd A starting from the largest value of N-Z, and going on in decreasing order.
 The middle panel
exhibits the same mass differences plotted against the order number, from 1 to 1654. 
It provides a univalued function,
whose Fourier transform can be calculated. The squared amplitudes are presented 
in Fig. \ref{fourier-l}.
As it was the case for the cuts for given N, Z, A ot T$_z$, the fully ordered data 
Fourier amplitudes are completely dominated by a few low frequencies. 
It is also remarkable that for ln(f) $>$ -3 the frequency distribution 
clearly resembles a power law.

\begin{figure}[h]  
  \begin{center}
    \psfig{file=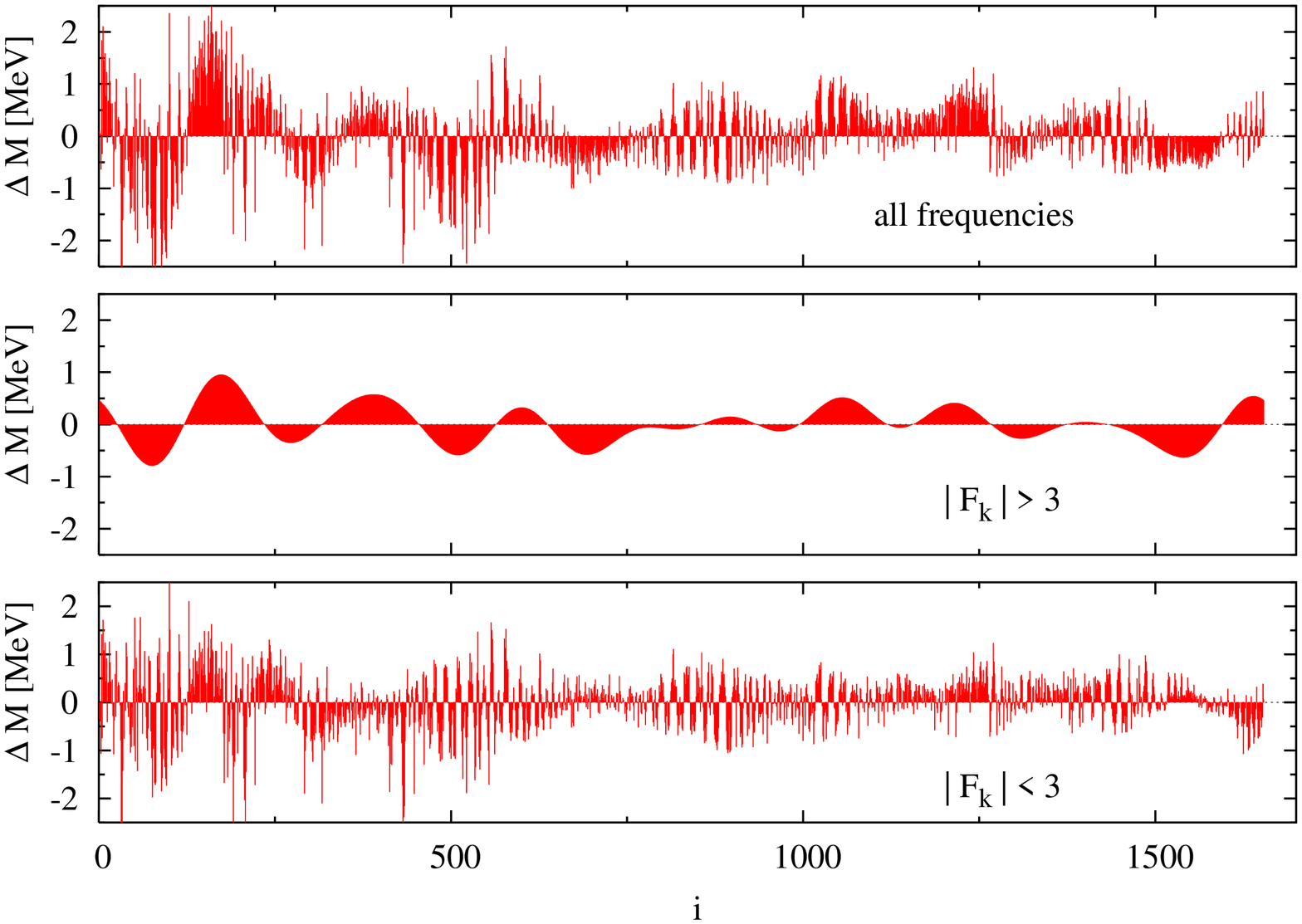, width=10.0cm,angle=270}
  \end{center}
%\vskip -3.cm
\caption{Mass differences plotted as an ordered list (top).
Mass error function recovered when only those
Fourier amplitudes larger than 3 are considered (middle) , and the mass error function
which remains when the larger Fourier amplitudes are removed (bottom).} 
\label{difmas-trunc}
\end{figure}
Removing the five frequencies with the largest Fourier components ($|F_k| > 3$)
generates a noisier pattern. 
Figure \ref{difmas-trunc} shows the mass differences plotted as function 
of the order index, and the results of leaving only the frequencies with amplitudes
$|F_k|$ larger (or smaller) than 3. There only five frequencies with Fourier amplitudes
whose absolute value is larger than 3. From them the dominant frequency is the one with period
$\Delta i = 207$. The regular pattern generated by these five frequencies in displayed
in the middle panel of Fig. \ref{difmas-trunc}. The remnant, shown in the bottom panel,
is clearly closer to white noise, while some bumps remain.

From the analysis presented in this section, we can conclude that there are conspicuous
correlations in the FRDM mass differences, whose periodic character
is clearly exhibited by the Fourier analysis of different cuts. 
It is worthwhile to study the slope of the squared amplitudes against the frequencies
(i.e. the power spectra) in a log-log plot, which in a random matrix context was found to offer a signature of quantum chaos \cite{Rel02}.  
This is of great  interest in itself and will be analyzed  in a separate publication \cite{Vel03b}.

\section{Removing the regularities}

Having established that there are patent regularities in the differences between the
masses calculated using the FRDM \cite{Moll95} and the measured ones, we will proceed to eliminate
them in a simple way, by removing the two frequencies which contribute the most
to the mass errors as functions of N and Z. We introduce an amplitude and a phase for
each frequency, having a total of six parameters for protons and six for neutrons.
The functions which minimize the errors  separately for protons and neutrons are:
\begin{eqnarray}
\Delta_{1} (N) = .30 \,\,\mbox{\rm sin}(2 \pi .012 N + .284) + 
		 .20 \,\,\mbox{\rm sin}(2 \pi .047 N + 1.06)\\
\Delta_{1} (Z) = .23 \,\,\mbox{\rm sin}(2 \pi .020 Z - 0.83) + 
		 .10 \,\,\mbox{\rm sin}(2 \pi .053 Z + 3.74) 
\end{eqnarray}

In a fit including all nuclei, we found
\begin{eqnarray} 
\Delta_{2} (N) = .21 \,\,\mbox{\rm sin}(2 \pi .011 N + 0.86) + 
		 .34 \,\,\mbox{\rm sin}(2 \pi .049 N + 0.17)  
\nonumber \\
\Delta_{2} (Z) =  .14 \,\,\mbox{\rm sin}(2 \pi .025 Z - 0.74)
		 -.19 \,\,\mbox{\rm sin}(2 \pi .075 Z - 5.53) 
\nonumber \\
	 \Delta_{2} (N, Z) = \Delta_{2} (N) + \Delta_{2} (Z) ,
\end{eqnarray}
while including only nuclei with A $\ge$ 65 the best fit is
\begin{eqnarray}
\Delta_{3} (N) = .65 \,\,\mbox{\rm sin}(2 \pi .009 N + 2.03) + 
		 .19 \,\,\mbox{\rm sin}(2 \pi .047 N + 1.00)  
\nonumber \\
\Delta_{3} (Z) = 2.90 \,\,\mbox{\rm sin}(2 \pi .018 Z - 3.51) + 
		 2.61 \,\,\mbox{\rm sin}(2 \pi .019 Z - 0.74) 
\nonumber \\
	 \Delta_{3} (N, Z) = \Delta_{3} (N) + \Delta_{3} (Z) .
\end{eqnarray}
The $\Delta_{1}$ are functions only of N or Z, and were adjusted using six parameters, optimized
for the 1350 nuclei with A $\ge$ 65. $\Delta_{2}$ and $\Delta_{3}$ are obtained by including
at the same time the corrections in Z and N, for all the nuclei in the first case, and for those
 with A $\ge$ 65 in the second. The fitted frequencies are all between 0.01 and 0.05,
in agreement with the dominant Fourier frequencies found in the previous section.

\begin{figure}[h]  
  \begin{center}
    \psfig{file=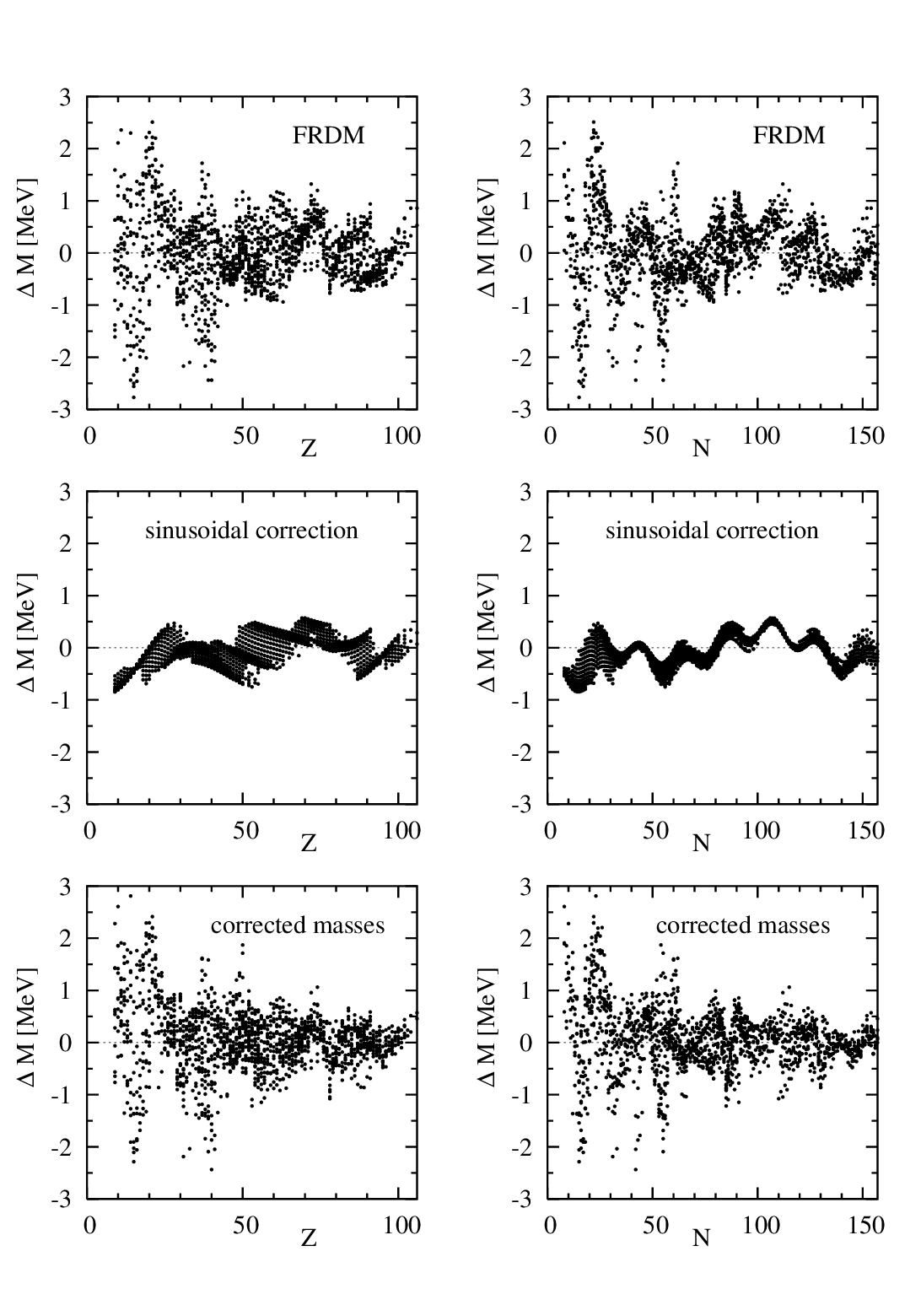, width=13.0cm}
  \end{center}
%\vskip -3.cm
\caption{Mass differences plotted as function of Z (left) and N (right),
for A $\ge$ 65.
The upper panels exhibit the FRDM results, the intermediate ones
the sinusoidal corrections, and the lower panels the corrected mass differences.}
\label{difsin}
\end{figure}

The effect of removing these sinusoidal components
as functions of N and Z is shown in Fig. \ref{difsin}, for those nuclei with masses larger than 65.
While the regular pattern is more apparent as a function of N, on the right hand side panels,
in both cases the `corrected' masses are more compact and compressed at smaller errors.

\begin{figure}[h]  
  \begin{center}
    \psfig{file=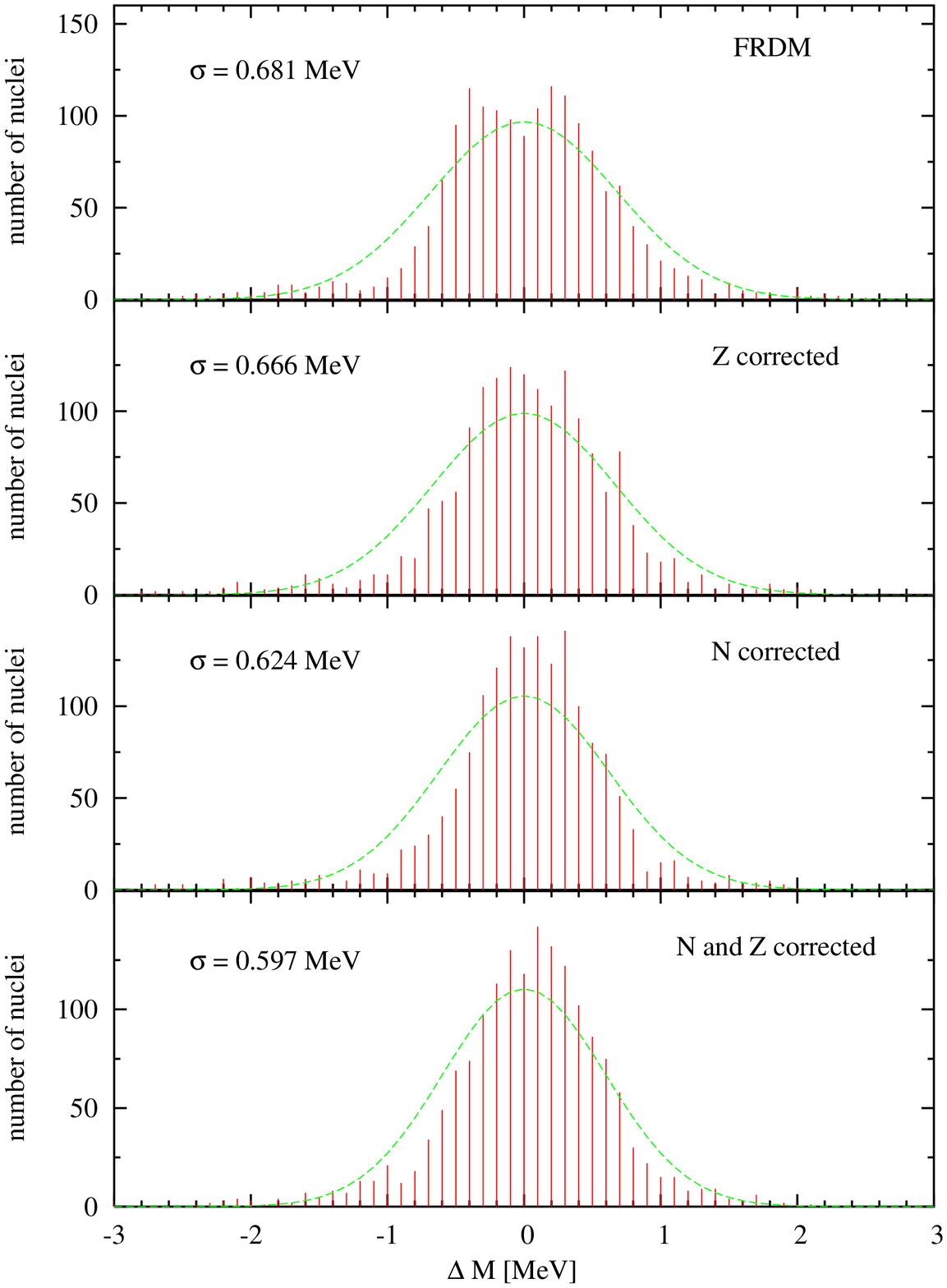, width=13.0cm}
  \end{center}
%\vskip -3.cm
\caption{Distribution of FRDM mass differences corrected 
with a sinusoidal function of Z, of N, and of Z and N.  }
\label{histo}
\end{figure}

\begin{figure}[h]  
  \begin{center}
    \psfig{file=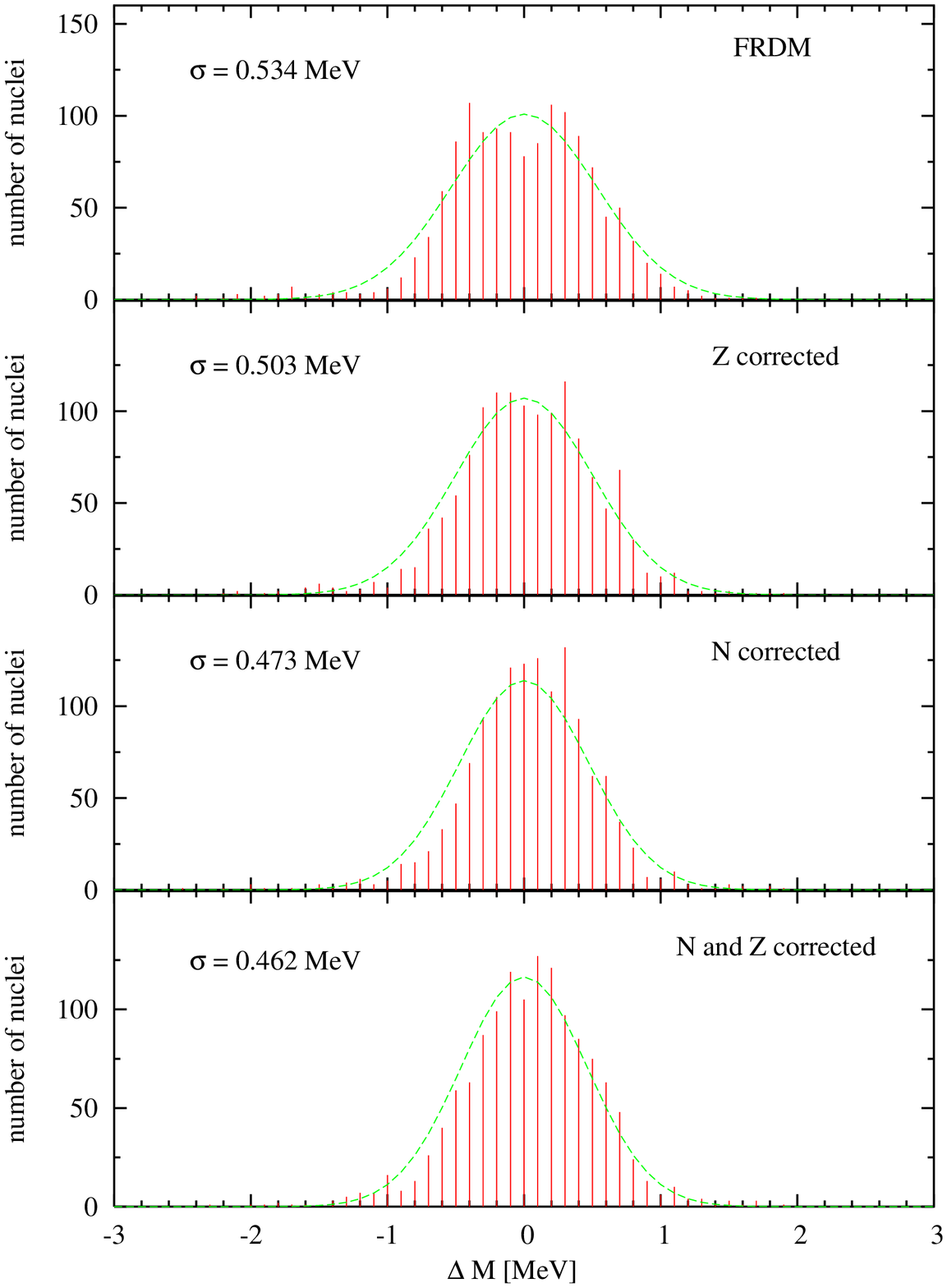, width=13.0cm}
  \end{center}
%\vskip -3.cm
\caption{Distribution of FRDM mass differences for A $\ge$ 65, corrected 
with a sinusoidal function of Z, of N, and of Z and N.  }
\label{histo-65}
\end{figure}

Obtaining a better fit of the known nuclear masses by the inclusion of six or twelve extra
variables is by no means surprising. However, the use of sinusoidal functions of N and Z
is strongly motivated by the data themselves. The fit does not only reduce the overall
error, but makes the double peak in the error distribution disappear, as shown in Fig.
\ref{histo} for all the nuclei, and in Fig. \ref{histo-65} for nuclei with A $\ge$ 65.
In both figures the error width reduction is evident, as well as the concentration
of the mass differences around zero, with a single peak. The continuous curve represents
a Gaussian curve with the same width and normalization. 
After removing the oscillatory components, the r.m.s. mass errors is reduced from
0.681 MeV to 0.597 MeV for all nuclei, and from 0.534 MeV to 0.462 MeV for nuclei with A $\ge$ 65.

\section{Final remarks}

In the present study we have shown that the differences between the
masses calculated using the FRDM of M\"oller at al \cite{Moll95} 
and the measured ones have a well defined
oscillatory component as function of N and Z, which can be removed with an
appropriate fit, significantly reducing the error width, and concentrating
the error distribution on a single peak around zero. 

The remaining correlations can only originate in the microscopic 
terms in the mass formula, which in the FRDM are evaluated using the Strutinsky method \cite{Bra72}.
Having shown that these correlations have a simple and clear dependence in 
the proton and neutron numbers, we are studying
the possible removal of these effects by a refinement of the Strutinsky method,
whose results will be reported elsewhere \cite{Vel03b}.

\section*{Acknowledgements}

Acknowledgements: Relevant comments by R. Bijker, O. Bohigas, J. Dukelsky, J. Flores, 
J.M. Gomez, P. Leboeuf, S. Pittel, A. Raga, P. van Isacker, H. Vucetich and A. Zuker 
are gratefully acknowledged.
This work was supported in part by Conacyt, M\'exico.

\end{document}